\documentclass[]{elsart}

\usepackage{harvard}

\usepackage{graphicx}

\usepackage{amssymb}

\begin{document}

\begin{frontmatter}



\title{Problems and Progress in Flare Fast Particle Diagnostics}
\author{John C. Brown and Eduard P. Kontar}
\ead{john@astro.gla.ac.uk, eduard@astro.gla.ac.uk}
\ead[url]{http://www.astro.gla.ac.uk}
\address{Department of Physics \& Astronomy, University of Glasgow, G12 8QQ, UK}



\begin{abstract}
Recent progress in the diagnosis of flare fast particles is
critically discussed with the main emphasis on high resolution
Hard X-Ray (HXR) data from RHESSI and coordinated data from other
instruments.  Spectacular new photon data findings are highlighted
as are advances in theoretical aspects of their use as fast
particle diagnostics, and some important comparisons made with
interplanetary particle data.  More specifically the following
topics are addressed

(a)  RHESSI data on HXR (electron) versus gamma-ray line (ion)
source locations.

(b) RHESSI hard X-ray source spatial structure in relation to
theoretical models and loop density structure.

(c) Energy budget of flare electrons and the Neupert effect.

(d) Spectral deconvolution methods including blind target testing
and results for RHESSI HXR spectra, including the reality and
implications of dips inferred in electron spectra

(e) The relation between flare in-situ and interplanetary particle
data.
\end{abstract}


\end{frontmatter}

\section{INTRODUCTION}

\label{INTRODUCTION} We were invited to present a critical review
of the present state of diagnostic methods for energetic particles
in flares in the light of recent progress. To deal with all energy
ranges of ions and electrons and all the numerous diagnostic
techniques (Table 1) used for them is impossible in a 30-minute
talk and a short report like this, and we have concentrated almost
exclusively on hard X-ray diagnostics of electrons but with
mention of some other regimes. From other talks at this meeting,
it was clear that we have not yet taken fully on board the more
definitive objective testing of models which state of the art data
now allow. For example strong possibilities now exist of exploring
quantitatively the  possible non-isothermality of ultra-hot
thermal flare plasmas by means of HXR spectra (Brown, 1974),
rather than sticking to an isothermal fit, with potentially major
implications for estimates of the flare electron energy budget.
So a critical review is timely, especially in the light of recent
high quality data (especially the Ramaty High Energy Solar
Spectroscopic Imager (RHESSI) (Lin et al., 2002)) which is truly
designed to help unravel ambiguities in data interpretation.

It is useful to start by commenting on what we mean by ``fast"
particles. Hard X-rays (HXR) are, broadly speaking, those at
energies above the atomic line regime - roughly 10 keV. Such
photons can very well be emitted profusely by a very hot ($T >>
10^7$K) flare plasma so in the HXR sense these locally Maxwellian
electrons are ``fast", particularly in the Maxwellian tail. What
we implicitly mean by ``fast" particle, however, is particles of
energy exceeding thermal ones. Thus, these particles are
essentially not in a Maxwellian and are far from equilibrium and
their distribution function can be arbitrary. Such particles are
truly ``non-thermal". It has to be noted, however, that the
relevant mean free path of the particles may not be the
collisional one - wave interactions can drive distributions toward
some local steady distribution possibly, though not necessarily,
Maxwellian. Thus the fact that a loop is longer than the mean free
path of a (say) $50$ keV particle does not preclude that particle
being part of a locally near-Maxwellian distribution. On the other
hand, seemingly non-Maxwellian distributions can be a sum of
locally Maxwellian distributions.

We should also comment on why fast particles are important in
flare modelling. While interpreting the HXR spectral diagnostic
alone remains rather ambiguous, when combined with spatial and
temporal HXR data and data at other wavelengths the data seem to
be broadly consistent with a large fraction of flare impulsive
phase power being in electrons of $\geq 20$ keV and ions in the
$0.1$~-~$1$ MeV range. Fast particles may thus be vital in flare
energy transport in that phase (There are clear indications that
pre-heating and gradual phase heating must be by other
mechanisms). Secondly, dissipation of $100$ Gauss worth of
magnetic energy in a coronal plasma of density $10^{10}$cm$^{-3}$
delivers a mean energy of $25$ keV per particle. Such particles
have collisional mean free paths vastly larger than current sheet
thickness. Consequently reconnection theory cannot be wholly
credible if it treats the plasma as a fluid (MHD) and ignores
particle kinetics and/or the presence of waves.

\begin{table}
\centering
\begin{tabular}{l}
  \hline
  {\bf Electromagnetic radiation from particles:} \\
  \hline
  X-ray Bremsstrahlung from electrons,\\
Plasma waves, gyrosynchrotron,\\ free-free radio emission,\\
nuclear and annihilation $\gamma$-ray lines,\\
pion decay $\gamma$-ray component,\\
atomic collision diagnostics, nonthermal ionisation, H$\alpha$ impact polarization\\
    \hline
     {\bf Interplanetary Particles:}\\
    \hline
  electrons, ions, neutrons \\
\end{tabular}
\caption{The various particle diagnostics involved in flare
studies}
\label{table1}
\end{table}

Remote radiation measurements in principle comprise the set of
Stokes Intensities $I_{O,Q,U,V}(\lambda, {\bf \Omega}, t)$ as
functions of wavelength $\lambda$, line of sight direction ${\bf
\Omega}$, and time $t$. The spatial information on $f_{e,i}({\bf
p},{\bf r},t)$ inferable is limited by the line of sight
projection/integration in each pixel (${\bf \Omega}$) and by the
fact that many physically important scales (sheet thickness,
gyroradii, Debye length) are well below any spatial resolution
currently, or ever likely to be, attainable.

\section{RESULTS FROM RHESSI AND RELATED DATA}
\subsection{Structure of Paper}

The last decade or so has brought a vast wealth of new flare solar
activity data (GRO Compton, YohKoh, SOHO, TRACE, RHESSI, WIND,
CORONAS-F) over numerous energy ranges. Of these the most
pertinent to progress in fast particle diagnostics in flares are
the remote sensing high resolution HXR imaging spectrometry of
RHESSI and the interplanetary data from WIND (cf. papers in Lin
and Krucker at this meeting). In this Section we summarise some of
the most exciting imaging results from RHESSI. In Section 3 we
discuss in more detail the analysis of RHESSI spectra and in
Section 4 we touch briefly on other data.

\subsection{HXR versus Gamma-Ray Source Location.}

\begin{figure}
  \begin{center}
  \includegraphics[width=76mm]{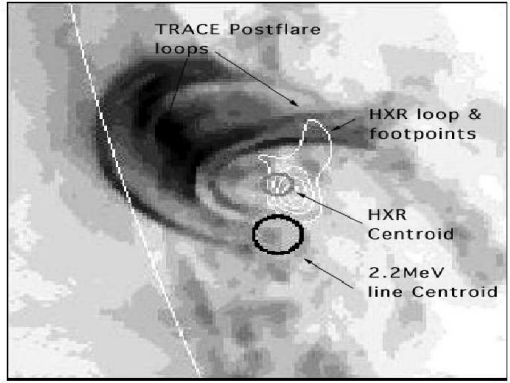}\\
  \end{center}
  \caption{Hard X-ray emission versus 2.2 MeV centroid location
  for July 23,2002 flare from Hurford et al., (2003).}
  \label{hurford}
\end{figure}

Gamma-ray events detected by RHESSI have been few in number but
offer some tantalising results for particle acceleration, some
contrasting with expectations. In particular the July 23 2003 2.2
MeV gamma-ray line image (Hurford et al., 2003) has a centroid
clearly separated from the centroid, or any part of, the HXR
(300-500 keV) image (Figure \ref{hurford}). This appears to
indicate that, in this event at least, MeV ions and deka-keV
electrons are accelerated and/or propagate in different parts of
the magnetic structure.  As yet the only quantitative
interpretation offered is that by Emslie, Miller and Brown (2004)
in which ions are preferentially accelerated in larger structures
than are electrons, or more generally in structures with longer
Alfven travel times (i.e. structures of greater size, greater
density, or lower field). The model can also yield the correct
order of accelerated fluxes and spectra but only insofar as a
suitable wave power is assumed.

\subsection{RHESSI HXR Images - Morphology, spectral structure and
evolution}

Fully reliable methods for reconstruction of spectrometric images
from RHESSI data are still under development but a variety of
important new results have already emerged (Emslie et al., 2003).
Though some HXR images show considerably more complexity than the
canonical ``two bright footpoints and faint coronal source", the
majority of RHESSI images, of sources large enough to be resolved,
do conform to that stereotype, at least approximately. Indeed, in
high resolution spectrometric images at progressively higher
energies (Aschwanden et al., 2002) show source separation from
soft looptop to hard footpoints in line with the Brown (1971)
thick target picture (The higher energy electrons penetrate deeper
into solar atmosphere and thus produce higher energy X-ray
emission in the region of higher density). The spectral index
difference between footpoints can be roughly understood in terms
of different column depth (Emslie et al., 2003). Aschwanden et al.
(2002) have proposed that the loop density structure implied by
these data, on the assumption of collisional transport, can be
used as an atmospheric density probe.

\begin{figure}
  \begin{center}
  \includegraphics[width=76mm]{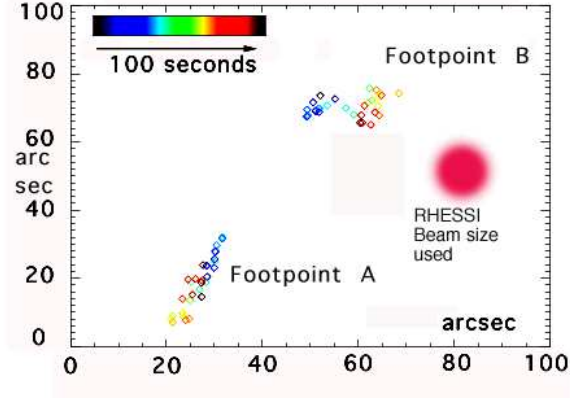}\\
  \end{center}
  \caption{Hard X-ray footpoint motion from Fletcher and Hudson (2002).}
  \label{fletcher_hudson}
\end{figure}

Fletcher and Hudson (2002) made a detailed study of HXR footpoint
motion, arguing that it rules out a single monolithic loop
structure throughout the event and suggesting that the source may
instead comprise a progressively activated sequence of very small
sources indicating the instantaneous bundle of field lines along
which electrons are being accelerated (Figure
\ref{fletcher_hudson}). Since the HXR flux defines the total
electron injection rate, this bundle cannot be too small ($>>$
current sheet thickness) or there would not be enough electrons
available even if all of them were accelerated.

\begin{figure}
  \begin{center}
  \includegraphics[width=76mm]{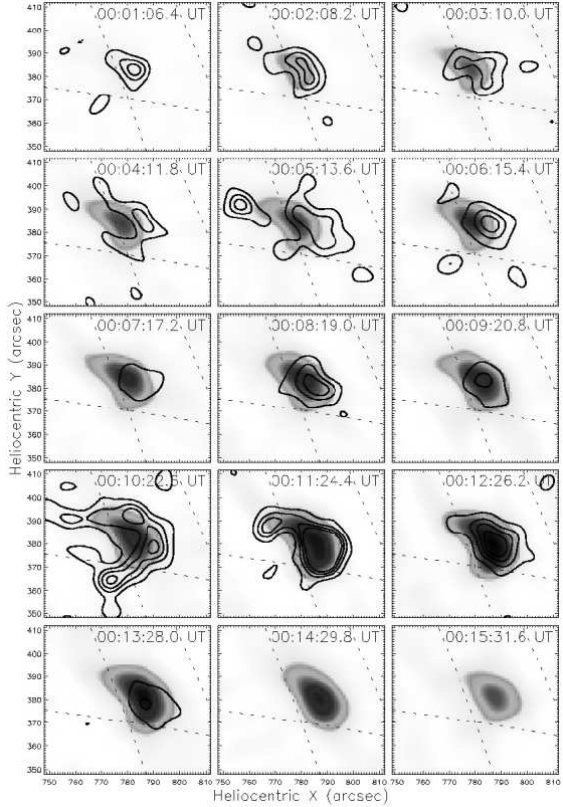}\\
  \end{center}
  \caption{April 14 flare. RHESSI images from Veronig et al. (2004).}
  \label{veronig}
\end{figure}

Dense thick target loop sources have been reported (Kosugi et al.,
1994; Veronig and Brown, 2004) in which there are essentially no
HXR footpoints, the entire loop emitting in both hard and soft XRs
(Figure \ref{veronig}). The high SXR loop emission measure
indicates a loop density high enough to stop  all but the highest
energy ($\geq 50$ keV) electrons. Such a scenario had in fact been
hinted at earlier (Kosugi et al, 1994) for a YohKoh event.

Veronig et al. (2004) have studied the evolution of loop densities
and temperatures and of HXR thick target beam parameters to test
the physics of the Neupert effect (Neupert, 1969), interpreted
purely as fast electron heating of loops.  They find that
including energy loss processes and comparing beam/plasma power
gives a generally poorer cross-correlation than the raw Neupert
HXR flux and time derivative of the SXR flux. They discuss
possible interpretations of this paradox in terms of variable low
energy cut-off, and of unresolved spatial structure including
possible sequential activation of small field line bundles.

Kane and Hurford (2003) have reported a number of sustained
coronal HXR sources of surprisingly large brightness and altitude
but as yet no physical interpretation has been offered.  These
pose tantalising questions as to what field structure can
accelerate and contain fast electrons in the corona.

While more work requires to be done, there are indications
(Schmahl and Hurford, 2002) that RHESSI images also contain
information on the photospheric albedo patch  around primary
sources, with the possibility of source height inference (Brown,
van Beek and McClymont 1975).

\section{HXR SPECTRAL INVERSIONS AND SOURCE ELECTRON SPECTRA}

\begin{figure}
  \begin{center}
  \includegraphics[width=76mm]{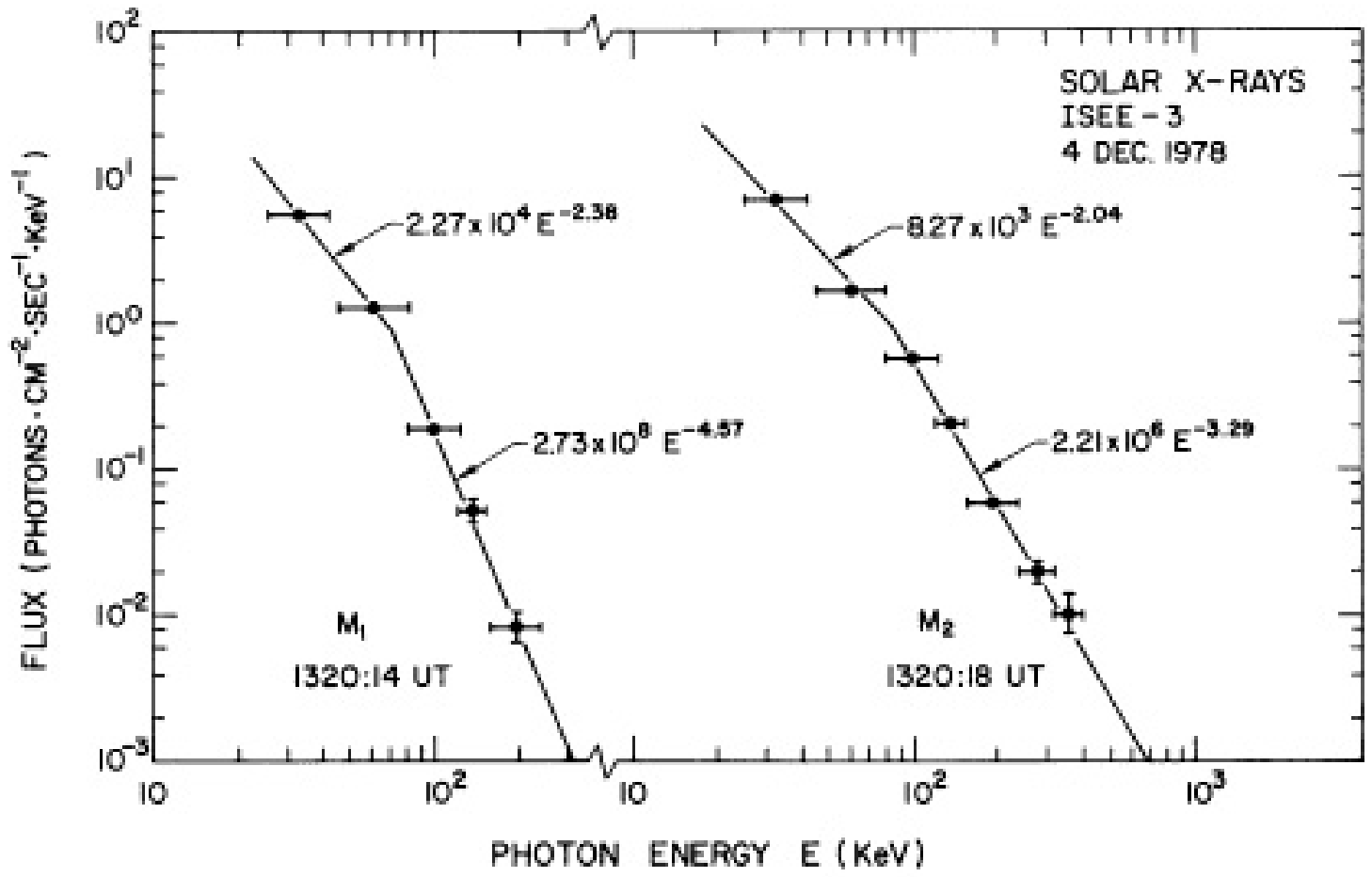}
  \includegraphics[width=76mm]{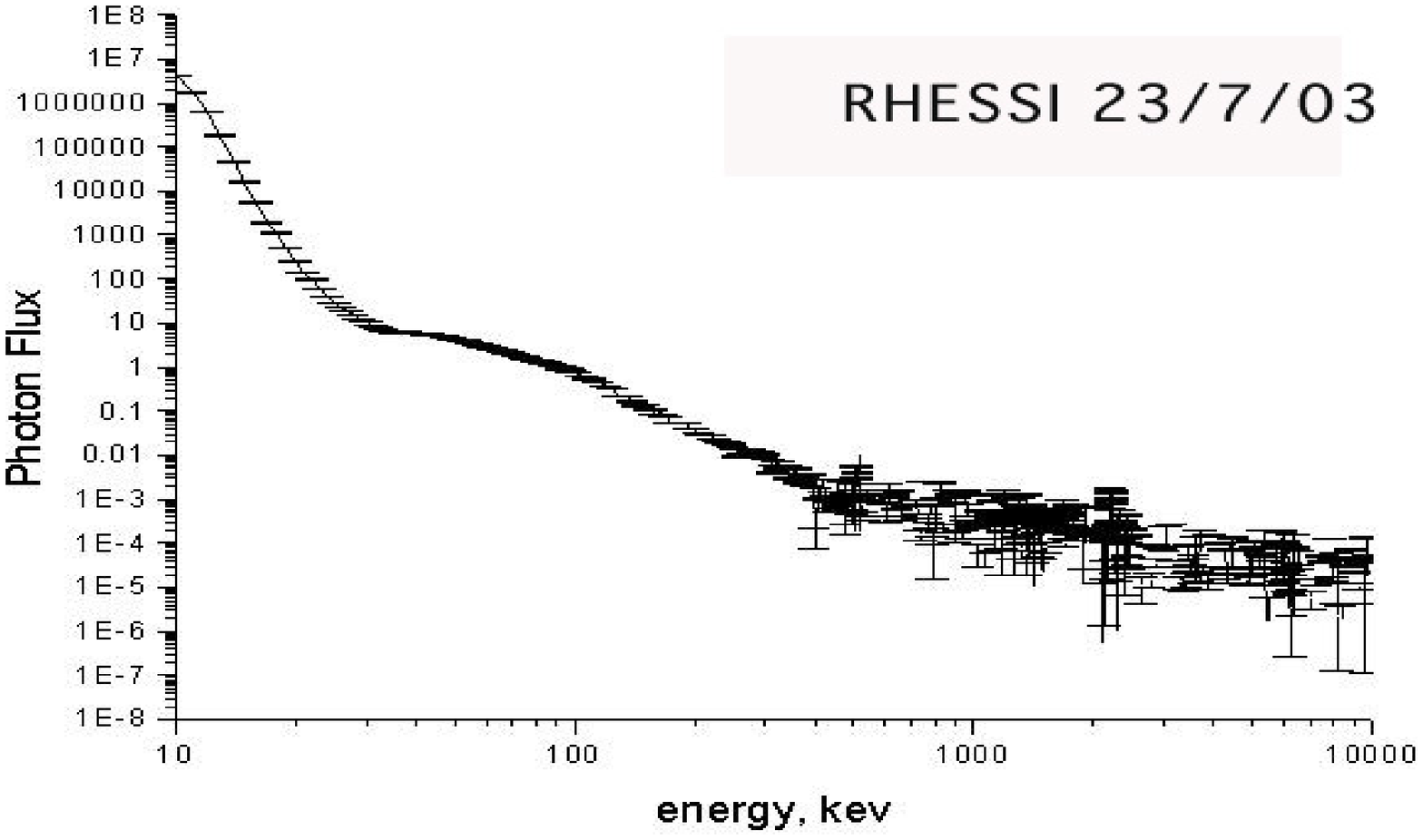}
  \end{center}
  \caption{X-ray spectrum from 80s (Kane, Benz and Treumann, 1982)
  and RHESSI spectrum}
  \label{hxr}
\end{figure}

\begin{figure}
  \begin{center}
  \includegraphics[width=76mm]{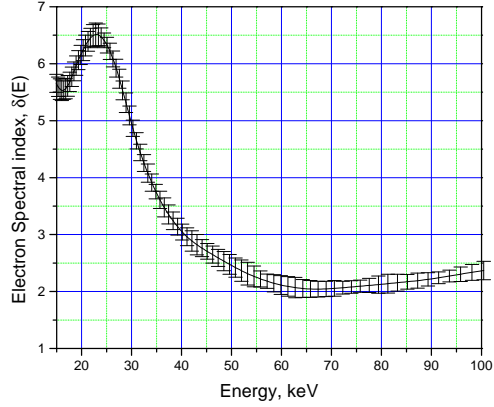}\\
  \end{center}
  \caption{Variation of local electron spectral index
  for Aug 21, 2002 M-class flare from Kontar and MacKinnon (2005).}
  \label{kontar05}
\end{figure}

\begin{figure}
  \begin{center}
  \includegraphics[width=76mm]{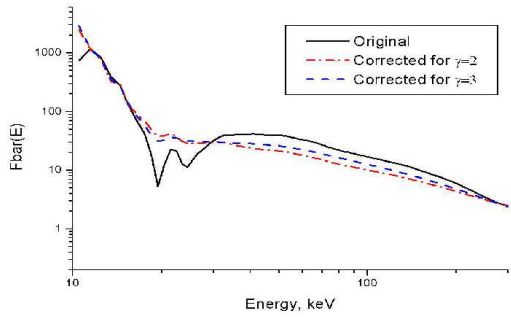}\\
  \end{center}
  \caption{Albedo correction and ``dip" in mean electron
  spectrum for August 20, 2002 solar flare from Kontar et al. (2004).}
  \label{kontar}
\end{figure}

\begin{figure}
  \begin{center}
  \includegraphics[width=76mm]{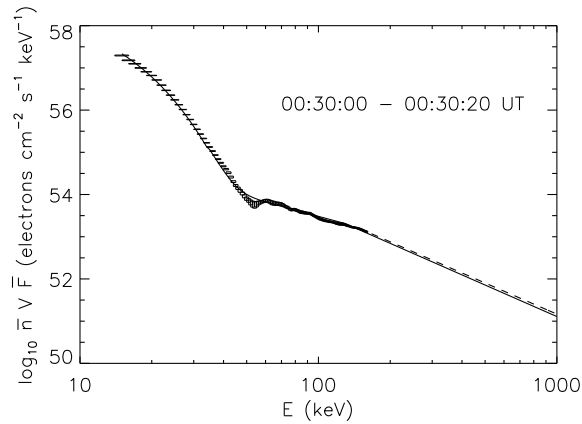}\\
  \end{center}
  \caption{Mean electron spectrum obtained for July 23,
  2002 flare from Piana et al. (2003).}
  \label{piana}
\end{figure}

As far as whole source HXR spectra are concerned (Figure
\ref{hxr}) shows how far RHESSI has advanced over typical previous
data, with photon spectral resolution of $\sim 1$ keV. These
enable for the first time (apart from Lin and Schwartz 1984) the
systematic inference of source electron spectra following Brown
(1971) and subsequent refinements and numerical implementations
(Johns and Lin (1992); Thompson et al. (1992); Piana et al.
(2003); Kontar et al. (2004)) to allow for regularised noise
suppression. Such inversions are now possible with such precision
as to yield the mean source electron spectrum $\bar F (E)$ and
local electron spectral index $\delta (E)$ as detailed functions
of electron energy $E$ (Figure \ref{kontar05}). RHESSI data
inversions of this kind are revealing a range of very interesting
electron spectral features including variable high energy cut-offs
(Kontar et al., 2004) and especially ``dips" in the spectrum where
$\delta (E)$ becomes very small or even negative (Figure
\ref{kontar})(Kontar and Brown, 2004). If such dips are proven to
exist in the primary HXR spectrum they rule out a purely
collisional thick target model in which the source electron
spectrum cannot have $\delta (E)<-1$ (Kontar and Brown, 2004). Two
possibilities that might make these inferred dips spurious are
detector pulse pile-up (Smith et al., 2002) and albedo
contributions (Alexander and Brown, 2002; Kontar, MacKinnon and
Brown, 2004). Work so far appears to rule out pile-up but shows
that albedo can create a spurious dip but at around $40$ keV. In
at least one case, shown in Figure \ref{piana}, the dip is around
$50$ keV and so may be real, though primary source directivity has
yet to be folded into the analysis. A genuine dip could be the
first direct inference of a low energy break in the electron
spectrum, crucial to the electron energy budget (Brown, 1971).


\begin{figure}
  \begin{center}
  \includegraphics[width=76mm]{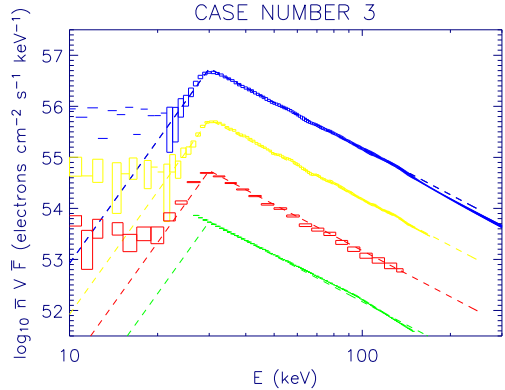}\\
  \end{center}
  \caption{The results of the inversions using various methods. Zero order
  regularization, first order regularization, regularization by coarse
  binning and forward fitting, respectively from top to bottom. The dash line shows
  the true solution. Successive curves have been scaled by 10  to render them visible.}
  \label{test}
\end{figure}

Given how vital the correct electron spectral shape is to testing
models, Brown et al. (2004) are carrying out systematic tests of
the reliabilities of different spectral inversion algorithms.  One
example of such a test is shown in Fig \ref{test}, which contains
the (blind) target spectrum and the results of three distinct
inversion algorithms, plus a forward best fit. Among the notable
conclusions are that all the inversion algorithms are good at
recovering dips and bumps, but that they do very badly in regimes
where the electron flux is small (since high energy electrons
swamp the photon data). These results are solely for an isotropic
cross-section. Generally, results are quite sensitive to the exact
form of the cross-section and so to the anisotropy of the electron
distribution (Massone et al., 2004). This situation is not as
discouraging as might first be thought. Massone (2004) have shown
that, at least in principle, bremsstrahlung spectra could contain
some information on both the angular and energy distribution of
the course electrons, analogously to the case of gyrosynchrotron
spectra (Fleischmann and Melnikov, 2003 ).

\begin{figure}
  \begin{center}
  \includegraphics[width=76mm]{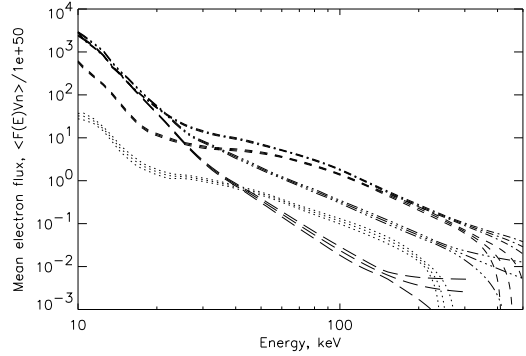}\\
  \end{center}
  \caption{High energy cut-off change for Feb 26, 2002 flare
  from Kontar et al. (2004b).}
  \label{kontar04}
\end{figure}

These ``mean source electron spectra" (Brown, Emslie and Kontar,
2003) are source model independent. Recently developed algorithms
for inferring mean source electron spectra (Kontar et al., 2004a)
show substantial variation of the spectral shape of the electron
spectrum as flares evolve (Kontar et al., 2004b). Application of
this technique to a flare on February 26, 2002 has shown that the
maximum accelerated electron energy rises and falls with time
after the peak of the event, concurrent with a growing low-energy
thermal component of the hard X-ray emission (Figure
\ref{kontar04}) (Kontar et al., 2004b).

\begin{figure}
  \begin{center}
  \includegraphics[width=76mm]{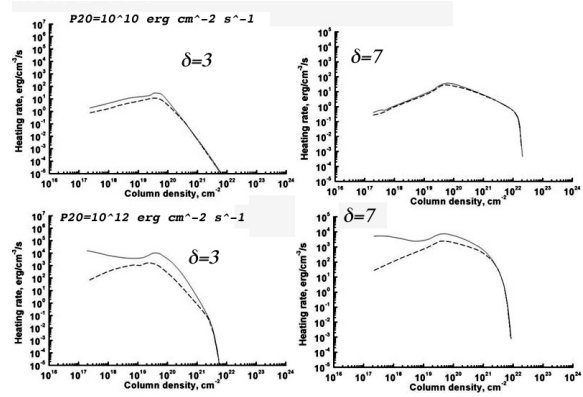}\\
  \end{center}
  \caption{Results of the simulations of beam propagation
  from Zharkova and Gordovskyy (2003).}
  \label{zharkova}
\end{figure}

Assuming propagation is dominated by collisions, one can infer
injected (accelerated) electron spectra.  To infer these
``injection" electron spectrum creating them is more uncertain
than finding mean spectra, requiring second deconvolution (Brown
and Emslie, 1988; Kontar et al., 2004b) and model assumptions such
as target ionisation structure (Kontar et al., 2003),
non-collisional energy losses (Zharkova and Gordovskyy, 2003)
(Figure \ref{zharkova}) and magnetic effects (e.g. mirroring) an
electron propagation.

Tests for pure thermality of the source spectrum (i.e.
superposition of Maxwellian) are also being developed but require
high order data derivatives (Brown and Emslie, 1988).

\section{INTERPLANETARY PARTICLES}

These have been discussed extensively by others at this meeting
and here we mention only a couple of points in relation to
remotely sensed data. The simultaneous operation of RHESSI, and of
TRACE, SOHO, GRO and KORONAS, with the interplanetary particle and
plasma probes aboard the WIND spacecraft is yielding many new
insights.  In particular the multi-made movies generated by
Krucker (2003) showing RHESSI SXR and HXR image evolution
superposed on TRACE images of near simultaneous EUV ``jets" formed
and associated with Type III bursts and interplanetary electrons
give clues to where the acceleration action is. For example, the
simultaneous upward and downward electron propagation places the
acceleration region in between.

The outward propagating electron streams are clearly visible via
their electromagnetic emission (Vilmer et al., 2003). WIND allows
us to follow these electron streams below ionospheric frequency
cut-off ($\sim 8$MHz) down to the local space plasma frequency
near the Earth orbit ($20$ kHz). For low energy electrons $\leq
50$ keV collective effects are crucial, since freely streaming
electrons build up unstable distribution functions. A recent
self-consistent approach (Melnik, 1995) shows that the generation
of Langmuir waves at the front of the stream, and absorption at
the back, lead to low spatial dispersion of electrons (Kontar et
al., 1998). These collective effects allow electrons to propagate
without substantial energy loss and are a source of the high level
of plasma turbulence required for Type III emission (Melnik and
Kontar, 2000).

\begin{figure}
  \begin{center}
  \includegraphics[width=76mm]{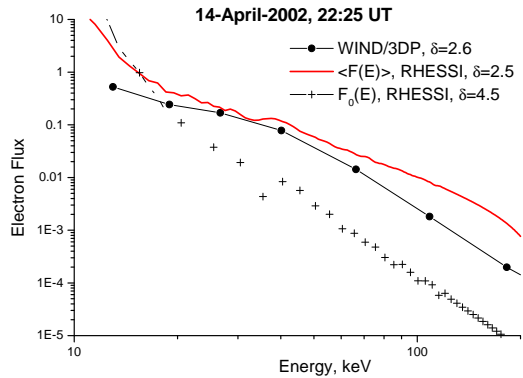}\\
  \end{center}
  \caption{Comparison of X-ray emitting electron spectrum and
  in-situ electrons from Krucker et al (2004).}
  \label{krucker}
\end{figure}

 Krucker et al (2003) have studied the spectral indices ($\delta
_{IP}$) of interplanetary (IP) electrons in relation to that at
the flare site implied by different models of the HXR source. They
find the fascinating result (Krucker, Kontar and Lin, 2004) that
$\delta _{IP}$ are much closer to the flare $\delta$ for
accelerated electrons if the electrons produce their HXRs in a
thin target rather than a collisionally thick one ($\delta_
{THICK} = \delta_ {THIN} + 2$) (Figure \ref{krucker}). This is
strange, and correspondingly important.  The possibility that HXRs
are purely thin target is hard to reconcile with HXR footpoints
and with the very large number of electrons it requires. An
alternative explanation is that collective effects act on the beam
(Haydock et al., 2001) to redistribute electron energies giving an
effective energy loss cross-section which is constant with energy
instead of $~1/E^2$ for collisions alone. Wave-wave interaction
brings additional complications and are sensitive to local plasma
inhomogeneities (Kontar and Pecseli, 2002)

\section{CONCLUSIONS}
    The interpretation of remotely sensed data at all wavelengths
continues to be riddled with ambiguities but the recent spate of
high resolution data from RHESSI and the numerous coordinated
observations from other instruments is truly starting to break
down these barriers to the understanding of fast particle
acceleration and propagation in flares.

{\bf Acknowledgements} We are thankful to A.G. Emslie, A. M.
Massone, S. Krucker, M. Piana, A. Veronig for valuable
discussions. We acknowledge the support of a PPARC Grant, a NATO
Collaboration Grant and the University of Alabama in Huntsville.

\end{document}